\documentclass[
    ,final            % use final for the camera ready runs
  ]
  {aipproc}

\layoutstyle{8x11single}

\usepackage{amsmath,amssymb,natbib,graphicx}
\usepackage{color}
\usepackage{ulem}

\begin{document}

\title{Afterglows of Mildly Relativistic Supernovae:\\Baryon Loaded Blastwaves}


\author{Sayan Chakraborti}{
  address={Department of Astronomy and Astrophysics, Tata Institute of Fundamental Research,\\
    1 Homi Bhabha Road, Mumbai 400 005, India}
}

\author{Alak Ray}{
  address={Department of Astronomy and Astrophysics, Tata Institute of Fundamental Research,\\
    1 Homi Bhabha Road, Mumbai 400 005, India}
}

\begin{abstract}
Relativistic supernovae have been discovered until recently only
through their association with long duration Gamma Ray Bursts (GRB).
As the ejecta mass is negligible in comparison to the swept up mass,
the blastwaves of such explosions are well described by the Blandford-McKee
(in the ultra relativistic regime) and Sedov-Taylor (in the non-relativistic
regime) solutions during their afterglows. However, the recent discovery of
the relativistic supernova SN 2009bb, without a detected GRB, has indicated
the possibility of highly baryon loaded mildly relativistic outflows which
remains in nearly free expansion phase during the radio afterglow. 
In this work, we consider the dynamics and emission from a massive,
relativistic shell, launched by a Central Engine Driven EXplosion (CEDEX),
decelerating adiabatically due to its collision with the pre-explosion
circumstellar wind profile of the progenitor. We show that
this model explains the observed radio evolution of the prototypical
SN 2009bb and demonstrate that SN 2009bb had a highly baryon loaded,
mildly relativistic outflow.
\end{abstract}

\maketitle

%%%%%%%%%%%%%%%%%%%%%%%%%%%%%%%%%%%%%%%%%%%%
%% MAINMATTER
%%%%%%%%%%%%%%%%%%%%%%%%%%%%%%%%%%%%%%%%%%%%

\section{Introduction}
Gamma-ray Bursts (GRBs) have long been recognized to require ultra relativistic bulk motion of matter particles (see
\citet{1999PhR...314..575P,2004RvMP...76.1143P} for reviews). GRB afterglows
are generated from 
the emission by relativistic shocks that result from slowing down of a relativistic
shell by the the medium surrounding the progenitor star that exploded.
In core collapse supernovae similar interaction of stellar material (ejecta) from an exploding star
with the circumstellar matter (CSM) results in non-relativistic shocks. 

Fluid dynamics of ultra-relativistic spherical blast waves mediated by strong 
shocks
has been treated by \citet{1976PhFl...19.1130B,1977MNRAS.180..343B}. They found a
similarity solution of
an explosion of a fixed amount of energy in a uniform medium.
%This includes an adiabatic
%blast wave and an impulsive injection of energy on a short timescale as well as an
%explosion where the total energy increases with time, suggesting that the blast wave
%has a continuous central power supply. Another important model
%considered by them is that of a blast wave propagating into a spherically symmetric wind.
On the other hand, \citet{1982ApJ...259..302C,1985Ap&SS.112..225N} described the initial nearly free expansion of a non-relativistic supernova
blast wave, interacting with the surrounding circumstellar medium. Once the
blast wave sweeps up more CSM material than its own rest mass, the self-similar
solutions of non-relativistic blast waves are described in the
Newtonian regime by the \citet{Sedov} \citet{vonNeumann} \citet{1950RSPSA.201..159T}
solution.

In this presentation we provide an analytic solution (see ApJ \citep{2011ApJ...729...57C} for details)
of the standard model of relativistic
hydrodynamics \citep[see e.g.][]{1999PhR...314..575P,1999ApJ...512..699C}
for an adiabatic blastwave.
Here, the exploding shell
decelerates due to inelastic collision with an external medium. The solution provided here is for an arbitrary Lorentz factor of the
expanding supernova shell. 
This solution which can handle a trans-relativistic outflow is motivated 
by the discovery of SN 2009bb, a type Ibc supernova without a detected GRB which
shows clear evidence of a mildly relativistic outflow powered by a central engine 
\citep{2010Natur.463..513S}. SN 2009bb-like
objects (Central Engine Driven Explosions, hereafter CEDEX \citep{2011ApJ...729...57C})
differ in another significant way from classical GRBs: our work shows that they are highly baryon
loaded explosions with non-negligible ejecta masses. 
The new analytic
blastwave solution here therefore generalizes the \citet{1976PhFl...19.1130B} result, 
in particular their impulsive, adiabatic blast wave in a
wind-like $\rho\propto r^{-2}$ CSM.

%The new class of relativistic supernovae without a detected GRB, e.g. SN 2009bb,
%relaxes two important and well-known constraints of the GRBs, namely the 
%\textit{compactness problem} and \textit{baryon contamination}. Thus highly baryon
%loaded mildly relativistic outflows can remain in the nearly free expansion
%phase during the radio afterglow. Therefore we consider the
%evolution of a massive relativistic shell launched by a CEDEX,
%experiencing collisional slowdown due to interaction with the pre-explosion
%circumstellar wind of the progenitor.

\section{Relativistic Blastwave Solution}
\label{blast}
We use the simple collisional model described by \citet{1999PhR...314..575P,1999ApJ...512..699C}
where the relativistic ejecta forms a shell which decelerates through
infinitesimal inelastic collisions with the circumstellar
wind profile. The initial conditions are characterized by the
rest frame mass $M_0$ of the shell launched by a CEDEX and its initial Lorentz
factor $\gamma_0$. 
%In contrast to the self similar solutions,
%which describe the evolution away from the boundaries of the independent
%variables \citep{1972AnRFM...4..285B}
%and the need for proper normalization to get the correct total energy,
%our set of initial conditions directly fixes the total
%energy as $E_0 = \gamma_0 M_0 c^2$.

%\subsection{Equations of Motion}
The shell slows down by collision
with the circumstellar  matter. The swept up circumstellar matter is given by
$m(R)$. Conservation of energy and momentum give us
\citep[see][]{1999ApJ...512..699C,1999PhR...314..575P},
\begin{equation}\label{dgamma}
\frac{d\gamma}{\gamma ^{2}-1} = -\frac{dm}{M} {\rm\;\;\; and \;\;\;}
dE=c^2(\gamma -1)dm ,
\end{equation}
%and
%\begin{equation}
%dE=c^2(\gamma -1)dm ,
%\label{dE}
%\end{equation}
respectively, where $dE$ is the kinetic energy converted into thermal energy, 
that is the energy in random motions as opposed to bulk flow, by
the infinitesimal collision.

For a circumstellar medium set up by a steady wind, where we expect a profile with
$\rho\propto r^{-2}$, we have $m(R)=AR$ where $A$ is the mass swept up by a sphere
per unit radial distance. $A\equiv\dot{M}/v_{wind}$ can be set up by a steady
mass loss rate of $\dot{M}$ with a velocity of $v_{wind}$ from the pre-explosion
CEDEX progenitor, possibly a Wolf Rayet star.
%Note that our definition of $A$ is equivalent to $4\pi q$
%in Equation (3 and 5) of \citet{1982ApJ...259..302C}.
Integrating the right hand side
and solving for $\gamma>1$ this equation we have
\begin{equation}
\gamma=\frac{\gamma_0 M_0 + A R}{\sqrt{M_0^2+2 A \gamma_0 R M_0 + A^2 R^2}} ,
\label{gammaR}
\end{equation}
which gives the evolution of $\gamma$ as a function of $R$.
%This is similar to Equation 8 of \citet{1999ApJ...512..699C} for
%a adiabatic spherical blastwave in a wind like ($\rho\propto r^{-2}$) CSM profile.
The amount of kinetic
energy converted into thermal energy when the shell reaches a particular $R$
can be obtained by integrating Equation (\ref{dgamma}) after substituting for $\gamma$
from Equation (\ref{gammaR}) and $dm=AdR$, to get
\begin{align}\label{ER}
E&=c^2 \left(-M_0-A R
%\right. \\ \nonumber
%&\left.
+\sqrt{M_0^2+2 A \gamma_0 R M_0+A^2 R^2}\right).
\end{align}

%\subsection{Evolution in Observer's Time}
The evolution of $R$ and $\gamma$ can be compared with observations once we have the
time in the observer's frame that corresponds to the computed $R$ and $\gamma$.
For emission along the line of sight from a blastwave with a constant $\gamma$
the commonly used expression \citep{1997ApJ...476..232M} is $t_{obs} =
R/(2 \gamma^2 c)$.
However, \citet{1997ApJ...489L..37S} has pointed out that for a decelerating
ultra-relativistic
blastwave the correct $t_{obs}$ is given by the \textit{differential} equation
$d t_{obs} = dR/(2 \gamma^2 c)$.
We substitute $\gamma$ from Equation (\ref{gammaR}) and integrate both sides to get
the exact expression
\begin{equation}
t_{obs}=\frac{R (M_0+A \gamma_0 R)}{2 c \gamma_0 (\gamma_0 M_0+A R)} .
\end{equation}
Note that, this reduces to the \cite{1997ApJ...476..232M} expression only in the
case of nearly free expansion and deviates as the shell decelerates.
In the rest of the work we use $t$ to indicate the time $t_{obs}$ in the observer's
frame. Inverting this equation and choosing the physically relevant \textit{growing branch},
gives us the analytical time evolution of the line of sight blastwave radius, as
\begin{align}\label{R}
R&=\frac{1}{2 A \gamma_0} \times \left(-M_0+2 A c \gamma_0 t
%\right. \\ \nonumber
%&\left.
+\sqrt{8 A c M_0 t \gamma_0^3+(M_0-2 A c \gamma_0 t)^2}\right) ,
\end{align}
in the ultra-relativistic regime.
This can now be substituted into Equations (\ref{gammaR} and \ref{ER}) to get the
time evolution of the Lorentz factor $\gamma$ and the thermal
energy $E$ \citep{2011ApJ...729...57C}.
This completes the solution for the blastwave time evolution,
parametrized by the values for $\gamma_0$, $M_0$ and $A$.

\begin{figure}
%\epsscale{1.1}
%\plotone{B_R.eps}
\includegraphics[width=0.45\columnwidth]{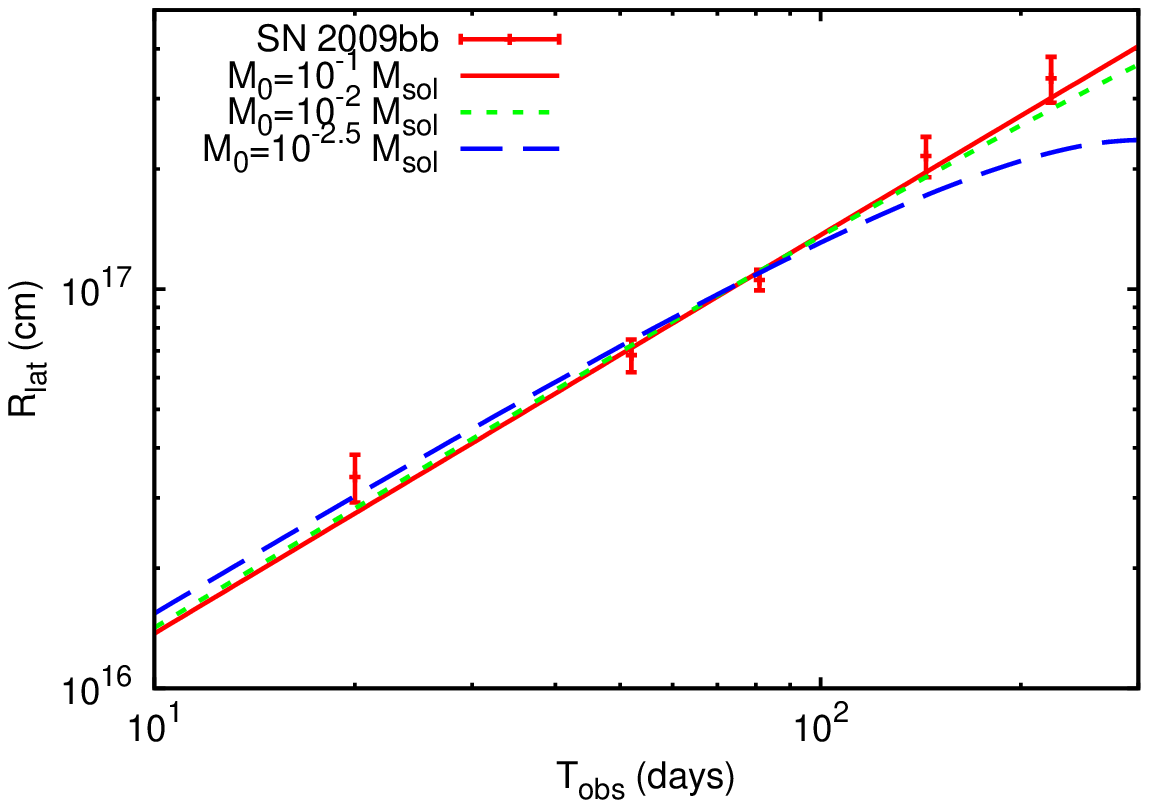}
\includegraphics[width=0.45\columnwidth]{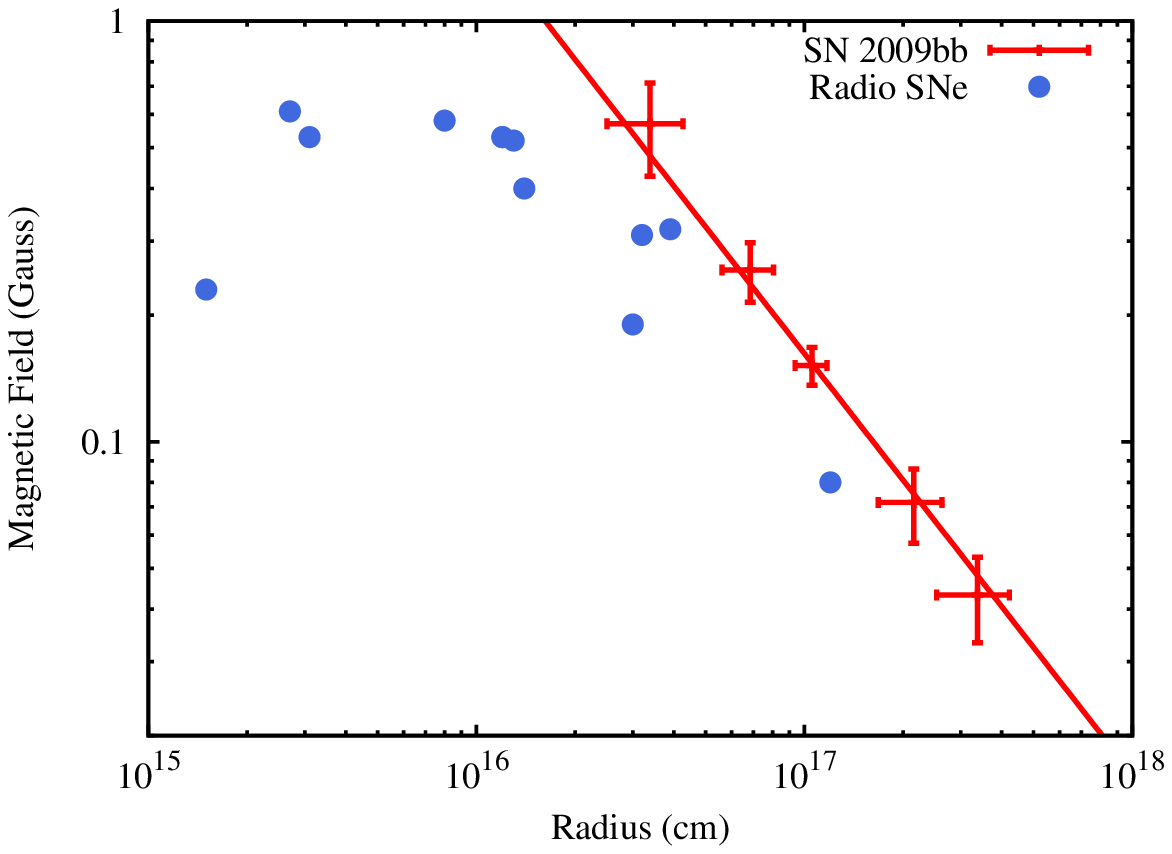}
\caption{\textbf{A:} Evolution of the blast wave radius $R_{lat}$, determined from SSA fit
to observed radio spectrum, as a function of the observer frame time $t_{obs}$.
The evolution is consistent with nearly free expansion and observations
require $M_0\gtrsim10^{-2.5}M_\odot$.
%\label{R_t}
%}
%\caption{
\textbf{B:} Magnetic field as a function of blast wave radius, as determined from
SSA fits. Blue dots represent size and magnetic field of radio supernovae from
\citet{1998ApJ...499..810C} at peak radio luminosity. Red crosses (with $3\sigma$
error-bars) give the size and magnetic field of SN 2009bb at different epochs, from
spectral SSA fits. Red line gives the best $B\propto R^{-1}$ (Equation \ref{B})
fit.\label{BR}}
\end{figure}

\section{Blastwave Energetics}
\label{energy}
\citet{2011ApJ...729...57C}
use the blastwave solution developed in the previous section to
predict the radio evolution of a CEDEX with a relativistic blastwave slowing
down due to circumstellar interaction.
For the prototypical SN 2009bb, the blastwave was only mildly relativistic
at the time of the observed radio afterglow. In the absence of a significant
relativistic beaming, the observer would receive emission from the entire
shell of apparent lateral extent $R_{lat}$ at a time $t_{obs}$ given by
$dt_{obs} = dR_{lat}/(\beta \gamma c)$.
%which is valid even in the mildly relativistic regime.
%Because of its mildly relativistic outflow, all expressions derived for
%SN 2009bb, use the above Equation rather than the\citet{1997ApJ...489L..37S}
%expression which
%is applicable in the ultra-relativistic case. In either case radio observations
%of SN 2009bb measure essentially the transverse $R_{lat}$ not the line of sight $R$.
Integrating term by term gives us the time evolution of the lateral
radius as $R_{lat}=c \sqrt{\gamma_0^2-1} t-O\left(t^2\right)$.
%\begin{align}\label{RLS}
%R_{lat}&=c \sqrt{\gamma_0^2-1} t
%-\frac{2 \left(A c^2 \gamma_0^3 \sqrt{\gamma_0^2-1}\right) t^2}{M_0}+O\left(t^3\right) .
%\end{align}
The thermal energy available when the shell has moved out to a radius $R$
is given exactly by Equation (\ref{ER}), however it is again convenient to
look at its Taylor expansion, $E=A c^2 (\gamma_0-1) R-O\left(R^2\right)$.
%\begin{align}
%E&=A c^2 (\gamma_0-1) R
%-\frac{\left(A^2 c^2 \left(\gamma_0^2-1\right)\right) R^2}{2 M_0}+O\left(R^3\right) .
%\end{align}

%\subsection{Electron Acceleration}
%\citet{1998ApJ...497L..17S} give the minimum Lorentz factor $\gamma_m$ of the
%shock accelerated electrons as
%\begin{equation}
%\label{gamma_m}
%\gamma_m=\epsilon_e\left(\frac {p-2} {p-1}\right) \frac {m_p} {m_e} \gamma.
%\end{equation}
%At the mildly relativistic velocities seen in SN 2009bb, the peak synchrotron
%frequency of the lowest energy electrons are likely to be below the synchrotron
%self absorption frequency. This explains the $\nu^{5/2}$ low frequency behavior
%of the spectrum. Hence,
We consider a electron distribution with an energy spectrum
$N_0 E^{-p}dE$, which we assume for simplicity to be extending from $\gamma_m m_ec^2$
to infinity, filling a fraction $f$ of the spherical volume of radius $R$.
%\begin{equation}
%E_e=\frac{4 f (\gamma_m m_ec^2)^{1-p} N_0 \pi  R^3}{3 (p-1)}.
%\end{equation}
If a fraction $\epsilon_e\equiv E_e/E$ of the available thermal energy goes into
accelerating these electrons, then for the leading order expansion of $E$ in $R$
the normalization of the electron distribution is given by
\begin{equation}\label{N0}
 N_0\simeq\frac{3 A c^2 \epsilon_e (\gamma_0-1) (\gamma_m m_ec^2)^2}{2 f \pi  R^2}, 
\end{equation}
for $p=3$, as inferred from the optically thin radio spectrum of
SN 2009bb \citep{2010Natur.463..513S}.

We consider a magnetic field of characteristic strength $B$ filling
the same volume.
If a fraction $\epsilon_B\equiv E_B/E$ goes into the magnetic energy density,
then the characteristic magnetic field is given by
\begin{equation}\label{B}
 B\simeq \frac{c}{R} \sqrt{\frac{6 A \epsilon_B (\gamma_0-1)}{f}}
\end{equation}
This explains the observed $B\propto R^{-1}$ behavior (Figure \ref{BR}) seen
in SN 2009bb. These observations strengthens the case for a nearly
constant $\epsilon_B$.

Note that the highest energy to which a cosmic ray proton can be accelerated is determined
by the $B \; R$ product \citep{1984ARA&A..22..425H,2005PhST..121..147W}.
We have argued elsewhere \citep{2010arXiv1012.0850C} that the mildly relativistic
CEDEX are ideal for accelerating nuclei to the highest energies
to explain the post GZK cosmic rays.

\section{Extracting Blast Wave Parameters from Radio Observations}
\label{inversiontv}
The inverse problem is that of determining the initial bulk Lorentz factor
specified by $\gamma_0$, progenitor mass loss rate given by $A$ or $\dot{M}$
and the initial ejecta mass $M_0$, from the radio observations. 
The bulk Lorentz factor may be determined from the radio observations using
to get the simplified expression for $\gamma_0$ as \citep{2011ApJ...729...57C},
\footnotesize
\begin{equation}
\gamma_0^2 \simeq 1+0.225\times
\left(\frac{\epsilon_B}{\epsilon_e}\right)^{2/19}
\left(\frac{f}{0.5}\right)^{-2/19} 
\left(\frac{t_{obs}}{20 {\rm~days}}\right)^{-2}
\left(\frac{\nu_p}{10 {\rm~GHz}}\right)^{-2}
\left(\frac{F_{\nu p}}{20 {\rm~mJy}}\right)^{18/19}
\left(\frac{D}{40 {\rm~Mpc}}\right)^{36/19}.
\label{gamma0TV}
\end{equation}
\normalsize
The result is insensitive to the equipartition parameter
$\alpha\equiv\epsilon_e/\epsilon_B$ and filling fraction $f$. This may be
used to reliably determine the initial bulk Lorentz factor of a radio detected
CEDEX in the mildly relativistic, nearly free expansion phase (like SN
2009bb).

A simplified expression for $A\equiv\dot{M}/v_{wind}$, the circumstellar
density profile,
set up by the mass loss from the progenitor has been derived by
\citet{2011ApJ...729...57C}. This gives us the approximate expression for the
mass
loss rate as
\begin{align}\nonumber
\dot{M} & \simeq 3.0\times10^{-6}
\left(\frac{\epsilon_B}{0.33}\right)^{-11/19}
\left(\frac{\epsilon_e}{0.33}\right)^{-8/19}
\left(\frac{f}{0.5}\right)^{11/19}
\left(\frac{v_{wind}}{10^3 {\rm~km s}^{-1}}\right)^{1} \\
&\times \left(\frac{t_{obs}}{20 {\rm~days}}\right)^{2}
\left(\frac{\nu_p}{10 {\rm~GHz}}\right)^{2}
\left(\frac{F_{\nu p}}{20 {\rm~mJy}}\right)^{-4/19}
\left(\frac{D}{40 {\rm~Mpc}}\right)^{-8/19}
{\rm~M_\odot yr^{-1}}.
\label{mdot0TV}
\end{align}
This approximate expression indicates the dependence of the inferred mass loss rate
on the observational parameters, and makes an error of only $\lesssim10\%$ in case of
SN 2009bb, when compared to our exact expression.
%Given the uncertainties in the
%observations, we recommend the use of this expression to get an estimate of the
%mass loss rate from a CEDEX progenitor.
Note that, this expression has similar
scaling relations as Equation (23) of \citet{2006ApJ...651..381C}. Hence, the
mass loss rate of SN 2009bb as determined using that equation by
\citet{2010Natur.463..513S} remains approximately correct.

The initial ejecta rest mass $M_0$ cannot be estimated from radio observations
in the nearly free expansion phase. It can
only be determined when the CEDEX ejecta slows down sufficiently due to interaction
with the circumstellar matter (Figure \ref{BR}). Thereafter, the initial ejecta
mass can be obtained using the timescale of slowdown
$t_{dec}$ and the already determined $A$ and $\gamma_0$. Nearly free expansion for a
particular period of time, can only put lower limits on the ejecta mass, as shown
in this work.

\section{Discussions}
\label{polao}
\citet{2011ApJ...729...57C} provide a solution of the relativistic hydrodynamics equations which is uniquely tuned to the
CEDEX class of objects like SN 2009bb. Because of the non-negligible initial ejecta
mass of such a CEDEX, objects like this would persist in the free expansion phase for
quite a long time into their afterglow. Sweeping up a mass equal to that of the
original ejecta would take considerable time, unless the mass loss scale in its
progenitor was very intense, i.e. it had a large $A$. We refer the reader to \citet{2011ApJ...729...57C} for a comparison of the CEDEX solution to the intermediate Blandford-McKee like solution and a final Snowplough phase.  
We also invert the dependence of the observed parameters on
$\gamma_0$ and $\dot{M}$ in terms
of the peak frequency, peak time and peak fluxes to interpret
the parameters of the explosion.

%%%%%%%%%%%%%%%%%%%%%%%%%%%%%%%%%%%%%%%%%%%%%%%%
%% BACKMATTER
%%%%%%%%%%%%%%%%%%%%%%%%%%%%%%%%%%%%%%%%%%%%%%%%

\begin{theacknowledgments}
We thank Alicia Soderberg, Abraham Loeb and Poonam Chandra for discussions on
mildly relativistic supernovae. We thank
Tsvi Piran and Charles Dermer for discussions at the Annapolis GRB2010 meeting.
\end{theacknowledgments}



\bibliographystyle{aipproc}   % if natbib is available

\bibliography{rel_sne}

\begin{thebibliography}{19}
\expandafter\ifx\csname natexlab\endcsname\relax\def\natexlab#1{#1}\fi
\providecommand{\enquote}[1]{``#1''}
\expandafter\ifx\csname url\endcsname\relax
  \def\url#1{\texttt{#1}}\fi
\expandafter\ifx\csname urlprefix\endcsname\relax\def\urlprefix{URL }\fi
\providecommand{\eprint}[2][]{\url{#2}}

\bibitem[{Piran}(1999)]{1999PhR...314..575P}
T.~{Piran}, \emph{\physrep} \textbf{314}, 575--667 (1999),
  \eprint{arXiv:astro-ph/9810256}.

\bibitem[{Piran}(2004)]{2004RvMP...76.1143P}
T.~{Piran}, \emph{Reviews of Modern Physics} \textbf{76}, 1143--1210 (2004),
  \eprint{arXiv:astro-ph/0405503}.

\bibitem[{Blandford} and {McKee}(1976)]{1976PhFl...19.1130B}
R.~D. {Blandford}, and C.~F. {McKee}, \emph{Physics of Fluids} \textbf{19},
  1130--1138 (1976).

\bibitem[{Blandford} and {McKee}(1977)]{1977MNRAS.180..343B}
R.~D. {Blandford}, and C.~F. {McKee}, \emph{\mnras} \textbf{180}, 343--371
  (1977).

\bibitem[{Chevalier}(1982)]{1982ApJ...259..302C}
R.~A. {Chevalier}, \emph{\apj} \textbf{259}, 302--310 (1982).

\bibitem[{Nadezhin}(1985)]{1985Ap&SS.112..225N}
D.~K. {Nadezhin}, \emph{\apss} \textbf{112}, 225--249 (1985).

\bibitem[{Sedov}(1946)]{Sedov}
L.~I. {Sedov}, \emph{Journal of Applied Mathematics and Mechanics} \textbf{10},
  241--250 (1946).

\bibitem[{von~Neumann}(1963)]{vonNeumann}
J.~{von~Neumann}, \emph{{The point source solution}}, Permagon Press, 1963, pp.
  219--237.

\bibitem[{Taylor}(1950)]{1950RSPSA.201..159T}
G.~{Taylor}, \emph{Royal Society of London Proceedings Series A} \textbf{201},
  159--174 (1950).

\bibitem[{Chakraborti} and {Ray}(2011)]{2011ApJ...729...57C}
S.~{Chakraborti}, and A.~{Ray}, \emph{\apj} \textbf{729}, 57 (2011),
  \eprint{arXiv:1011.2784}.

\bibitem[{Chiang} and {Dermer}(1999)]{1999ApJ...512..699C}
J.~{Chiang}, and C.~D. {Dermer}, \emph{\apj} \textbf{512}, 699--710 (1999),
  \eprint{arXiv:astro-ph/9803339}.

\bibitem[{Soderberg} et~al.(2010)]{2010Natur.463..513S}
A.~M. {Soderberg}, S.~{Chakraborti}, G.~{Pignata}, et~al., \emph{\nat}
  \textbf{463}, 513--515 (2010), \eprint{0908.2817}.

\bibitem[{Meszaros} and {Rees}(1997)]{1997ApJ...476..232M}
P.~{Meszaros}, and M.~J. {Rees}, \emph{\apj} \textbf{476}, 232 (1997),
  \eprint{arXiv:astro-ph/9606043}.

\bibitem[{Sari}(1997)]{1997ApJ...489L..37S}
R.~{Sari}, \emph{\apjl} \textbf{489}, L37--L40 (1997).

\bibitem[{Chevalier}(1998)]{1998ApJ...499..810C}
R.~A. {Chevalier}, \emph{\apj} \textbf{499}, 810--819 (1998).

\bibitem[{Hillas}(1984)]{1984ARA&A..22..425H}
A.~M. {Hillas}, \emph{\araa} \textbf{22}, 425--444 (1984).

\bibitem[{Waxman}(2005)]{2005PhST..121..147W}
E.~{Waxman}, \emph{Physica Scripta Volume T} \textbf{121}, 147--152 (2005),
  \eprint{arXiv:astro-ph/0502159}.

\bibitem[{Chakraborti} et~al.(2011)]{2010arXiv1012.0850C}
S.~{Chakraborti}, A.~{Ray}, A.~{Soderberg}, A.~{Loeb}, and P.~{Chandra},
  \emph{Nature Communications} \textbf{2}, 175 (2011),
  \eprint{arXiv:1012.0850}.

\bibitem[{Chevalier} and {Fransson}(2006)]{2006ApJ...651..381C}
R.~A. {Chevalier}, and C.~{Fransson}, \emph{\apj} \textbf{651}, 381--391
  (2006), \eprint{arXiv:astro-ph/0607196}.

\end{thebibliography}

\IfFileExists{\jobname.bbl}{}
 {\typeout{}
  \typeout{******************************************}
  \typeout{** Please run "bibtex \jobname" to optain}
  \typeout{** the bibliography and then re-run LaTeX}
  \typeout{** twice to fix the references!}
  \typeout{******************************************}
  \typeout{}
 }

\end{document}